\documentclass[a4paper,twocolumn, 11pt, unpublished]{quantumarticle}
\pdfoutput=1
\usepackage[utf8]{inputenc}
\usepackage[english]{babel}
\usepackage[T1]{fontenc}
\usepackage{amsmath}
\usepackage{hyperref}
\usepackage{multirow}
\usepackage{lipsum}
\usepackage[square,numbers]{natbib}
\usepackage{float}
\usepackage{mleftright} 
\usepackage[normalem]{ulem}
\usepackage{amsfonts}
\usepackage{float}
\usepackage{url}
\newcommand{\figref}[1]{\mbox{Fig.~\ref{#1}}}

\newcommand{\secref}[1]{\mbox{Sec.~\ref{#1}}}

\renewcommand{\eqref}[1]{\mbox{Eq.~(\ref{#1})}}

\newcommand{\bra}[1]{\mleft\langle #1 \mright |}
\newcommand{\ket}[1]{\mleft|#1 \mright \rangle}

\newcommand{\expec}[1]{\mleft\langle #1 \mright\rangle}

\newcommand{\be}{\begin{equation}}
\newcommand{\ee}{\end{equation}}
\newcommand{\bea}{\begin{eqnarray}}
\newcommand{\eea}{\end{eqnarray}}

\newtheorem{theorem}{Theorem}

\begin{document}

\title{Implicit differentiation of variational quantum algorithms}
\author{Shahnawaz Ahmed}
\affiliation{Department of Microtechnology and Nanoscience, Chalmers University of Technology, 412 96 Gothenburg, Sweden}
\affiliation{Xanadu, Toronto, ON, M5G 2C8, Canada}
\orcid{0000-0003-1145-7279}
\author{Nathan Killoran}
\affiliation{Xanadu, Toronto, ON, M5G 2C8, Canada}
\author{Juan Felipe Carrasquilla Álvarez}
\affiliation{Vector Institute for Artificial Intelligence, MaRS Centre, Toronto, ON, Canada M5G 1M1}
\affiliation{Department of Physics and Astronomy, University of Waterloo, Waterloo, Ontario, N2L 3G1, Canada}
\orcid{0000-0001-7263-3462}
\maketitle
\begin{abstract}
Several quantities important in condensed matter physics, quantum information, and quantum chemistry, as well as quantities required in meta-optimization of machine learning algorithms, can be expressed as gradients of implicitly defined functions of the parameters characterizing the system. Here, we show how to leverage implicit differentiation for gradient computation through variational quantum algorithms and explore applications in condensed matter physics, quantum machine learning, and quantum information. A function defined implicitly as the solution of a quantum algorithm, e.g., a variationally obtained ground- or steady-state, can be automatically differentiated using implicit differentiation while being agnostic to how the solution is computed. We apply this notion to the evaluation of physical quantities in condensed matter physics such as generalized susceptibilities studied through a variational quantum algorithm. Moreover, we develop two additional applications of implicit differentiation --- hyperparameter optimization in a quantum machine learning algorithm, and the variational construction of entangled quantum states based on a gradient-based maximization of a geometric measure of entanglement. Our work ties together several types of gradient calculations that can be computed using variational quantum circuits in a general way without relying on tedious analytic derivations, or approximate finite-difference methods.
\end{abstract}
\section{Introduction}
\begin{figure}
    \centering
    \includegraphics[width=\columnwidth]{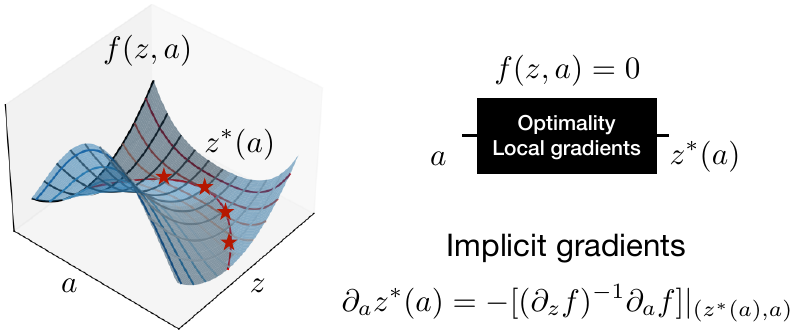}
    \caption{\label{fig:ift}Automatic differentiation through a black-box algorithm that implicitly defines a solution function $z^{*}(a)$ to some problem using an optimality condition $f(z, a) = 0$. Solution points (shown with the red stars) can be obtained using an iterative black-box solver without having access to the explicit solution. However, if we have access to local gradients ($\partial_z f, \partial_a f$) at the solution points, implicit differentiation can compute the gradient of the solution function $\partial_a z^{*}(a)$ without differentiating through the black box.}
\end{figure}
Variational quantum algorithms (VQAs) cast the solution of a computational problem as the classical optimization of the expectation value of an observable with respect to a state generated by powerful parametrized quantum circuits~\cite{Cerezo2021,RevModPhys.94.015004}. In light of their potential for enabling computational breakthroughs, VQAs have been developed for a wide range of applications encompassing most of the domains originally targeted by quantum computing, such as finding ground states of molecules, determining steady states of open quantum systems, simulating quantum dynamics, solving linear systems of equations, machine learning, combinatorial optimization, and many more~\cite{Cerezo2021, RevModPhys.94.015004}.

Variational quantum circuits can express highly-entangled complex quantum states beyond classical models such as deep Boltzmann machines and tensor networks with small bond dimensions~\cite{PhysRevResearch.2.033125}. The increased computational power of quantum states comes at a significant price related to the difficulty of estimating their properties, which are key to all scientific and technological applications of quantum computing~\cite{elbenRandomizedMeasurementToolbox2022}. An important family of such quantities can be written as gradients of functions computed on a quantum state. For instance, nuclear forces in quantum chemistry~\cite{obrienCalculatingEnergyDerivatives2019,pulayInitioCalculationForce1969,PhysRevResearch.2.013129}, permanent electric dipole~\cite{kassalQuantumAlgorithmMolecular2009}, static polarizability~\cite{kassalQuantumAlgorithmMolecular2009}, the static hyperpolarizabilities of various orders~\cite{kassalQuantumAlgorithmMolecular2009}, generalized susceptibilities~\cite{sandvikComputationalStudiesQuantum2010a}, fidelity susceptibilities~\cite{PhysRevE.76.022101}, and geometric tensors~\cite{PhysRevLett.99.095701}. Moreover, in the emerging area of quantum machine learning, analogous gradients may prove critical in the development of hyperparameter tuning, meta-learning, and architectural search techniques. Thus, automatizing gradient computations may expand the algorithmic flexibility of VQAs and ease the development of a wider variety of gradient computations by removing the tedious burden of computing and implementing these derivatives by hand.

A natural approach to tackle problems involving gradient computation within a VQA is automatic differentiation (AD). Recently, AD has seen significant development within the quantum information research community in part due to the availability of modern quantum software such as PennyLane~\cite{bergholmPennyLaneAutomaticDifferentiation2022} and TensorFlow Quantum~\cite{broughtonTensorFlowQuantumSoftware2021}, which provide tools to implement AD for functions evaluated on simulated and physical quantum devices. While it is already straightforward to calculate several important gradients in VQAs, most prominently energy gradients (with respect to the parameters in the circuits~\cite{PhysRevA.98.032309,wierichsGeneralParametershiftRules2022}, and with respect to parameters in the Hamiltonian through the Hellmann-Feynman theorem~\cite{obrienCalculatingEnergyDerivatives2019}), gradient computation of expectation values of general operators using standard AD techniques remains generally unavailable. These generic gradients may become computationally difficult since in principle they require keeping track of all intermediate optimization steps in a VQA. 

Here we develop an implicit differentiation strategy to automatically evaluate gradients and higher-order derivatives of expectation values of generic operators beyond the Hellman-Feynman theorem. Specifically, this implicit AD approach applies for the case where the operators, states, or gates making up the expectation value are themselves obtained as the solution of a complex optimization problem involving a quantum circuit, such as finding the ground state of a Hamiltonian. 

Our strategy combines differentiable programming of quantum computers~\cite{bergholmPennyLaneAutomaticDifferentiation2022} with modular implicit differentiation~\cite{blondelEfficientModularImplicit2022} as a tool to automatically evaluate implicit gradients through variational quantum algorithms. A sketch explaining the idea of implicit differentiation is given in \figref{fig:ift}, and \figref{fig:scheme} shows how this idea can be extended to VQAs. We present three applications of using this technique: computing generalized susceptibilities, hyperparameter tuning in a quantum machine learning algorithm, and obtaining a Bell state by variationally optimizing a geometric measure of entanglement. Our implementation integrates a classical modular implicit differentiation package (JAXopt~\cite{blondelEfficientModularImplicit2022}) with a quantum AD framework (PennyLane~\cite{bergholmPennyLaneAutomaticDifferentiation2022}), which allows us to effortlessly add implicit differentiation to VQAs.

The paper is organized as follows. In \secref{sec:implicit} we discuss the implicit function theorem and the derivations of gradient formulas for implicitly defined functions. In \secref{sec:applications} we discuss the prototypical example of differentiating through the ground state of a Hamiltonian to demonstrate how implicit differentiation can be leveraged through a ground-state-finding algorithm. Then, in \secref{sec:examples}, we show three applications of our approach. First, we show how generalized susceptibilities can be computed through a VQA that obtains the ground state of a parameterized Hamiltonian in \secref{sec:susceptibility}. Then we show two examples that go beyond differentiating ground-state solutions. In \secref{sec:hyperparameter} we demonstrate how hyperparameter optimization of a quantum machine learning algorithm is possible using implicit differentiation. Finally, in \secref{sec:entanglement} we generate a maximally entangled Bell state variationally by optimizing an entanglement witness. 
The data and code for all numerical experiments are available in~\cite{code}.

\section{Implicit differentiation}
\label{sec:implicit}
The main tool we use in our work is the implicit function theorem (IFT). Below we state the theorem in its simplest form for an analytic function~\cite{krantz2002implicit, Hormander1973}.

\begin{theorem}[IFT for analytic functions]
Let $f(z, a)$ be an analytic function of complex variables $(z, a)$ in a neighbourhood around the point $(z_0, a_0) \in \mathbb C^{m}\times\mathbb C^{n}$. If $f(z_0, a_0) = 0$ and $\partial_z f \ne 0$ at $(z_0, a_0)$, then there exists a unique analytic solution function $z^{*}(a)$ to $f(z, a) = 0$ in a neighbourhood $\mathcal N(a_0)$, given by $(z_0, a_0)$ with $z^{*}(a_0) = z_0$.
\end{theorem}
Since the solution function $z^{*}(a)$ is analytic, it can be complex-differentiable in the neighbourhood of points $(z^{*}(a), a)$ that satisfy
\begin{equation}
    \label{eq:optimality}
    f(z^{*}(a), a) = 0.
\end{equation}
This allows computing gradients of the solution function by differentiating \eqref{eq:optimality} w.r.t $a$ at ($z_0 = z^{*}(a_0), a_0$) 
\begin{gather}
\label{eq:implicit-formula}
    \partial_a f(z_0, a_0) + \partial_{z} f(z_0, a_0) \partial_{a} z^{*}(a) = 0 \nonumber \\
    \partial_{a} z^{*}(a) = - (\partial_{z} f(z_0, a_0) )^{-1}\partial_a f(z_0, a_0)
\end{gather}
The region $\mathcal N(a_0)$ that defines the analytic domain can be lower bounded as shown in~\cite{chang2003analytic}.

In practice, AD tools can be used to compute the gradients $\partial_a f$ and $\partial_z f$, which are vectors for a multivariate function. In fact $f$ can also be vector-valued such that $\partial_a f$ and $\partial_z f$ are Jacobians. These can be computed column-wise with vector-Jacobian products (VJPs) or row-wise using Jacobian-vector products (JVPs). VJPs—also called reverse-mode autodifferentiation or backpropagation—are efficient for a scalar-valued function, compared to JVPs, where we only have one output but several input parameters. VJPs make the calculation of implicit gradients easier, as we show below.

Mathematically, the VJP of a function $g:\mathbb R^n \to \mathbb R^m$ for input parameters $x \in \mathbb{R}^n$ is the operation
\begin{gather}
    \text{VJP}(g, x, v): (x, v) \to (v^{T}\partial g(x)),
\end{gather}
where the vector $v \in \mathbb{R}^m$ has the same dimension as the output of the function $g(x)$ and $\partial g:\mathbb R^{n}\to\mathbb R^{m}\times \mathbb R^{n}$ is the Jacobian matrix. It is easy to see that setting $v = [1, 1, ... ]$ gives us the gradients of the function for all $n$ parameters,
\[
v^{T}\partial g(x) = \begin{bmatrix}
  1, & 1, & \dots
\end{bmatrix}
\begin{bmatrix}
  \frac{\partial g_1}{\partial x_1} & 
    \hdots & 
    \frac{\partial g_1}{\partial x_n} \\[1ex] 
  \vdots & 
    \vdots & 
    \vdots \\[1ex]
  \frac{\partial g_m}{\partial x_1} & 
    \hdots & 
    \frac{\partial g_m}{\partial x_n}
\end{bmatrix}.
\]

We can write a VJP for the implicit function $z^{*}(a)$ from \eqref{eq:implicit-formula} as the operation
\begin{align}
    \text{VJP}(z^{*}, a, v): (a, v) \to & (v^{T}\partial_a z^{*}(a)) \nonumber \\
     = & (-v^{T}A^{-1}B),
\end{align}
assuming $A = \partial_z f(z_0, a_0)$ and $B = \partial_a f(z_0, a_0)$.

VJPs solve the difficulty of obtaining the inverse of the matrix $A$ (matrix inversion has a cubic scaling). Using VJPs we can avoid the cost of full inversion with a trick that only requires approximately solving the linear system~\cite{blondelEfficientModularImplicit2022}
\begin{equation}
    v = A^{T}u
\end{equation}
for a vector $u$ that has the same dimension as $a$. The approximate solution to this linear system using an iterative algorithm such as generalized minimal residual method or conjugate gradient~\cite{blondelEfficientModularImplicit2022} simplifies the inversion as
\begin{eqnarray}
    [A^{T}]^{-1} v &=& u \nonumber \\
    ([A^{T}]^{-1} v)^{T} &=& u^{T} \nonumber \\
    v^{T}A^{-1} &=& u^{T}.
\end{eqnarray}
We can then compute the VJP of the implicit function as
\begin{eqnarray}
    \text{VJP}(z^{*}, a, v): (a, v) \to (-u^{T}B), \nonumber \\
\end{eqnarray}
which gives $\partial_a z^{*}(a) = -u^{T}B$. 

Implicit differentiation can also be applied to fixed-point equations such as
\begin{equation}
    f(z, a) = z,
\end{equation}
by defining $h(z, a) = z - f(z, a)$, with $h(z, a)=0$, such that a solution map $z^{*}(a)$ satisfies
\begin{equation}
    h(z^{*}(a), a) = 0.
\end{equation}
A similar derivation for such fixed-point equations gives the gradients as
\begin{equation}
\label{eq:fixedpoint}
    \partial_a z^{*}(a) =  [\mathbf I - A]^{-1}B.
\end{equation}
In \eqref{eq:fixedpoint} the inversion can also be approximated with a similar trick as discussed before, where we now define
\begin{equation}
\label{eq:vdef}
    v = [\mathbb I - A^{T}]u
\end{equation}
such that we have
\begin{eqnarray}
\label{eq:fixedpointinv}
    v + A^{T}u  &=& u.
\end{eqnarray}
\eqref{eq:fixedpointinv} now defines a fixed-point function for $u$ that can be solved using any fixed-point solver. Then using \eqref{eq:vdef} we have
\begin{equation}
    u^{T} = v^{T}[\mathbb I - A]^{-1},
\end{equation}
and the VJP for the fixed point solution can be computed as
\begin{equation}
    \text{VJP}(z^{*}, a, v): (a, v) \to (-u^{T}B).
\end{equation}
The implicit gradient is again $\partial_a z^{*}(a) = -u^{T}B$.

Therefore, calculating implicit gradients efficiently requires computing VJPs of a function defining a fixed point equation or optimality condition at the solution point. The inverse term is approximated efficiently with a simple trick that requires solving a linear system of equations or iteratively finding a fixed point. Other approaches for calculating the inverse such as using a Neumann series approximation~\cite{lorraineOptimizingMillionsHyperparameters2020, Rajeswaran2019} are also possible.

\begin{figure}
    \centering
    \includegraphics[width=\columnwidth]{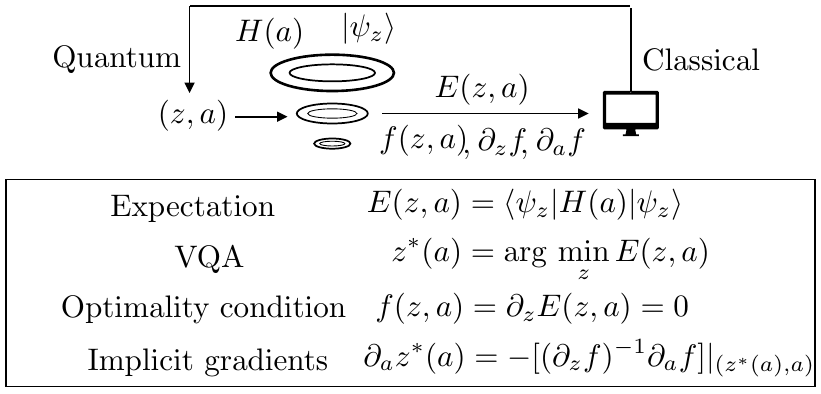}
    \caption{\label{fig:scheme}Implicit differentiation through variational quantum algorithms: A quantum computer generates a variational state $\ket{\psi_{z}}$ and computes expectation values of quantities such as the energy $E(z, a)$ of a parameterized Hamiltonian $H(a)$. A classical optimization provides a solution map $z^{*}(a)$ for variational parameters $z$,  minimizing the expectation value and therefore defining an optimality condition. We can then apply implicit differentiation to calculate gradients $\partial_a z^{*}(a)$ which would otherwise require unrolling all the steps in obtaining $z^{*}(a)$ using standard automatic differentiation (AD). In a realistic VQA, unrolling the optimization and storing intermediate variables may not be feasible for direct application of AD. However using the higher-order gradients of expectation values computed by the quantum computer at the solution we can compute $\partial_a z^{*}(a)$ using implicit differentiation.}
\end{figure}

Now, we can connect implicit differentiation to quantum algorithms, as illustrated in \figref{fig:scheme}.

\section{Implicit differentiation of variational quantum algorithms}
\label{sec:applications}

We can use implicit differentiation on variational quantum algorithms in several different ways, depending on which elements of the expectation value (e.g., the states or the observable) are parameterized. As a motivating example, consider a variational quantum algorithm whose objective function is defined through an operator $H(a)$, with parameters $a$, that codifies the solution to a computational problem. Here, $H(a)$ may represent a quantum Hamiltonian, the objective function of a quantum machine learning algorithm, an Ising Hamiltonian encoding a combinatorial optimization problem, etc. Variational quantum algorithms optimize the expected value of $H(a)$ over a quantum state generated by a parameterized quantum circuit, represented by a unitary matrix $U(z)$ endowed with parameters $z$. The unitary $U(z)$ acts on an input state $\ket{\psi_0}$, typically a product state, as $\ket{\psi_z} = U(z)\ket{\psi_0}$. A quantum computer provides a strategy to evaluate the expectation value of $H(a)$ on $\ket{\psi_z}$, 
\begin{equation}
    E(z, a) = \expec{\psi_z|H(a)|\psi_z},
\end{equation}
as well as its derivatives with respect to $z$.  A variational quantum algorithm minimizes $E(z,a)$ with respect to $z$, which gives optimal parameters
\begin{equation}
\label{opt}
    z^{*}(a) = \arg\,\min_{z} E(z, a).
\end{equation}
 Concretely, suppose we are interested in approximating the ground state of a quantum many-body Hamiltonian $H(a)$. The algorithm aims at minimizing the expectation value of the Hamiltonian over a variational quantum circuit to find an approximation to  its ground-state properties.
 
The solution $z^{*}(a)$ corresponds to a minimum of the function $E(z, a)$ which means that 
\begin{equation}
    \label{eq:quantum-func}
    f(z^{*}(a), a) = \partial_z E(z^{*}(a), a) = 0.
\end{equation}
The approximate ground-state solution of $H(a)$,   $\ket{\psi_{z^{*}(a)}}$, is therefore implicitly defined by $z^{*}(a)$ and depends on the Hamiltonian parameters $a$. We are interested in computing how a change in the Hamiltonian parameter $a$ affects the solution $z^{*}(a)$, and functions of the solution state.

Beyond the motivating example of a parameterized Hamiltonian, we can also construct implicit gradients for an arbitrary function that combines variational parameters $z$ and parameters of interest $a$, e.g., a loss function $\mathcal L(z, a)$ defined with a variational state $\psi_z$, where $a$ represents the strength of regularization. Similarly, we can consider $z$ and $a$ to be variational parameters for two different quantum circuits, as we show in our example in \secref{sec:entanglement} using a geometric definition of an entanglement measure using two variational quantum states. The example of taking gradients through a ground-state minimization task is perhaps most straightforward, and therefore we focus on it in more detail below.

The gradient of an expectation value defines a function $f(z, a) = \partial_a E(z, a)$ along with an optimality or fixed-point condition, i.e., $f(z, a) = 0$. In this case, the function $f(z, a)$ itself is the first-order gradient of the expectation value. Provided we can compute the Hessian of $E(z, a)$  w.r.t. parameters $z$, and its inverse, implicit differentiation gives a strategy to evaluate $\partial_a z^{*}(a)$. As we demonstrate below, $\partial_a z^{*}(a)$ can be used to compute other quantities that depend on the ground state. 

We first consider a function that computes the expectation value of an operator $A$ on the ground state, 
\begin{equation}
    \expec{A} = \expec{\psi_{z^*(a)}|A|\psi_{z^*(a)}}.
\end{equation}
If $A = H(a)$, i.e., we are computing the ground-state energy, the Hellman-Feynmann theorem~\cite{PhysRev.56.340} can be applied to obtain
\begin{equation}
    \partial_a \expec{A} = \expec{\psi_{z^*(a)}|\partial_a A|\psi_{z^*(a)}}.
\end{equation}
In quantum-chemistry problems such as molecular geometry optimization~\cite{pulayInitioCalculationForce1969}, the Hellman-Feynmann theorem simplifies the gradient computation as we only require expectation values of $\partial_a A$.
However, in a more general case, when the operator $A$ is not the Hamiltonian, the computation of $\partial_a \expec{A}$ requires implicit gradients
\begin{eqnarray}
\label{eq:gen-hellman}
    \partial_a \expec{A} &=& \partial_a z^*(a)\cdot \partial_z \expec{A} \nonumber \\
    &+& \expec{\psi_{z^*(a)}|\partial_a A|\psi_{z^*(a)}}. \nonumber\\
\end{eqnarray}

All the quantities required to compute \eqref{eq:gen-hellman} can be evaluated on a quantum computer for a large family of quantum circuits. Assuming the existence of a parameter shift-rule \cite{PhysRevA.98.032309, schuld2019evaluating, wierichsGeneralParametershiftRules2022} for $U(z)$, the term $\partial_z \expec{A}$ can be evaluated using a similar strategy used to compute gradients of the objective function of the variational quantum algorithm.  Likewise, the second term, which is analogous to the Hellman-Feynmann theorem, is evaluated by computing the expectation value of an operator $\partial_a A$ on the variationally obtained solution $\psi_{z^{*}(a)}$. Finally, the evaluation of $\partial_a z^*(a)$  (see Eq.~\ref{eq:implicit-formula}) requires access to $\partial_a f, \partial_z f$, which are  second-order derivatives of the expectation value $E(z,a)$, accessible on a quantum computer for large classes of variational circuits~\cite{PhysRevA.103.012405}.

Our implicit differentiation formula can therefore extend the capabilities of automatic differentiation tools such as PennyLane to compute gradients of a variationally-obtained solution state without relying on tedious analytic computations or numerical approximations. Further, the computation does not require keeping track how the solution is computed or unrolling all the steps in the optimization, which might be infeasible in a variational quantum computation.
\section{Examples}
In the following examples we show how implicit differentiation can be applied to calculate quantities such as generalized susceptibility, perform hyperparameter optimization in a quantum machine learning algorithm, and to generate multipartite entanglement using variational quantum circuits.
\label{sec:examples}
\subsection{Generalized susceptibility calculation}
\label{sec:susceptibility}
The response of an observable $A$ upon the introduction of a perturbation in a Hamiltonian $H(a)$ can be quantified through a generalized susceptibility. These types of linear response functions are written in terms of the gradient of the expectation value of the operator $A$ measured on the ground state $\psi_{\text{GS}}(a)$, 
\begin{equation}
    \partial_a \expec{A} = \partial_a \bra{\psi_{\text{GS}}(a)}A\ket{\psi_{\text{GS}}(a)},
\end{equation}
where $\psi_{\text{GS}}(a)$ represents the ground state. Here $a$ represents, e.g., the strength of a field $B$ in the Hamiltonian, and $A$ is the operator whose response to this perturbation we aim to compute. Examples of generalized susceptibilities important in condensed matter physics include the uniform magnetic susceptibility, or transport properties such as the superfluid stiffness, all of which are central to the characterization of strongly correlated phases of matter and their corresponding phase transitions~\cite{sandvikComputationalStudiesQuantum2010a}. 

The computation of susceptibilities and fidelity susceptibilities are typically carried out analytically for sufficiently tractable problems~\cite{PhysRevB.99.165130,PhysRevB.76.180403,kirillovExactSolutionHeisenberg1986}, using finite-difference methods through exact diagonalization and density matrix renormalization group~\cite{PhysRevB.99.165130,PhysRevA.87.043606}, through a costly computation of excited states using exact or variational techniques~\cite{sakaiQuantumSpinFluid2018}, or through quantum Monte Carlo simulations for Hamiltonians which do not have a sign problem~\cite{sandvikComputationalStudiesQuantum2010a}.
\begin{figure}
    \centering
    \includegraphics[width=\columnwidth]{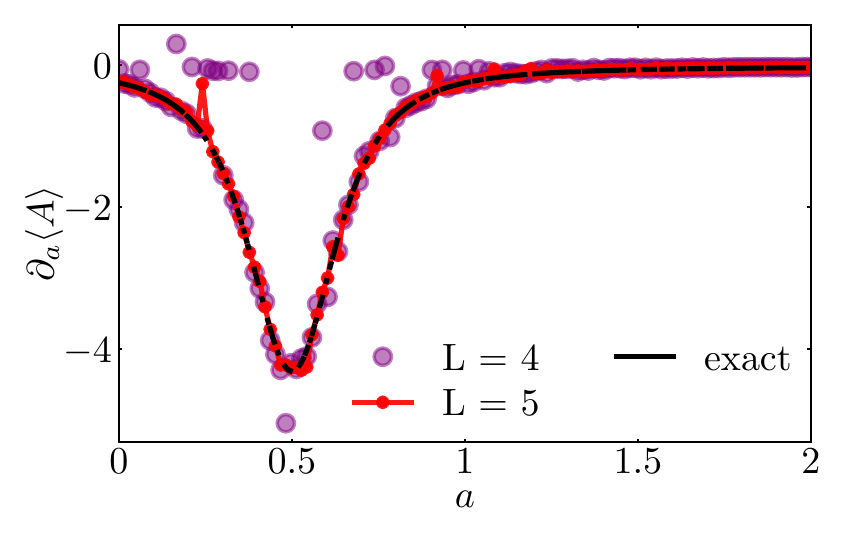}
    \caption{\label{fig:sus}Susceptibility computation for a $N = 5$ spin system. The exact computation finds the ground state of a parameterized Hamiltonian $H(a)$ and unrolls the eigendecomposition to compute gradients (susceptibility) $\partial_a \expec{A(a)}$ using AD. In the variational setting, we can approximate the ground state by minimizing the expectation value of $H(a)$ using a variational state $\vert\psi_z\rangle$. The variational state is generated by the two-design ansatz with $L$ layers. Once the ground-state solution $z^{*}(a)$ is found, implicit differentiation can compute the susceptibility efficiently without requiring to keep track of how the variational ground state is obtained.}
    
\end{figure}
Alternatively, we can compute susceptibilities in a variational quantum algorithm using implicit differentiation. Here, we consider a parameterized Hamiltonian for a linear spin-chain model in an external magnetic field given by
\begin{eqnarray}
    H(a) &=& -\sum_{i} \sigma^z_i \sigma^z_{i+1} - \gamma \sum_{i} \sigma^x_i - a \sum_i \sigma^z_i. \nonumber 
\end{eqnarray}
We aim at computing the response of the average magnetization $
    \partial_a \expec{A}$
to a change in a magnetic field along the $z$ direction, i.e., $A = \frac{1}{N}\sum_i \sigma^z_i$ and  $B=\sum_i \sigma^z_i$. Here $\sigma^{x, y, z}_{i}$  represent the Pauli spin operators at the site $i$. A small field  $\delta \sum_i \sigma^z_i$ is added to the Hamiltonian for numerical stability near $a\approx0$.

In a variational setting, a parameterized ground state $\ket{\psi_{z^*(a)}}$ can be obtained by optimizing the state $\ket{\psi_z}$ with respect to its variational parameters $z$. With the variationally-obtained ground state, by applying implicit differentiation we can compute the susceptibility using \eqref{eq:gen-hellman}, going beyond the Hellman-Feynmann theorem.

In our experiments, the variational ground state is obtained by running a standard VQE optimization. Our variational circuit is a simplified two-design ansatz consisting of Pauli-Y rotation gates and entangling controlled-Z gates. We can consider multiple layers for a more expressive ansatz. The parameters of the circuit are optimized with respect to the energy $E(z, a)$ using gradient descent.

In \figref{fig:sus} we show how implicit differentiation through a variationally-optimized ground state compares against an exact calculation of the susceptibility $ \partial_a \expec{A}$. The exact ground state $\ket{\psi_{\text{GS}(a)}}$ is obtained through eigendecomposition of $H(a)$, where we evaluate $ \partial_a \expec{A}$ by differentiating through the eigendecomposition~\cite{boeddekerComputationComplexvaluedGradients2019,PhysRevB.101.245139}. The gradient computation through eigendecomposition is implemented with the AD tool Jax~\cite{jax2018github} using the formulation in ~\cite{boeddekerComputationComplexvaluedGradients2019} (specifically Eq.~(4.60)). For a sufficiently expressive ansatz with $L = 5$ layers, we can closely match the susceptibility computed by eigendecomposition. However for $L = 4$ layers, the gradient computation significantly deviates from the exact values.

Compared to the eigendecomposition approach, which scales exponentially with system size, the variational approach may scale well in practice for larger system sizes in future quantum devices beyond systems that can be simulated classically. Provided that the variational algorithms find accurate representations of the ground state and the availability of parameter-shift rules, we anticipate that variational quantum algorithms paired with implicit differentiation will enable the computation of susceptibilities and other response functions in such systems. While we have derived the susceptibility formula in Eq.~(\ref{eq:gen-hellman}) motivated by their use in a quantum algorithm, we emphasize that  Eq.~(\ref{eq:gen-hellman}) can be applied to the simulation of response functions in classical variational techniques such as in neural network quantum states~\cite{carleoSolvingQuantumManybody2017} and density matrix renormalization group calculations~\cite{PhysRevLett.69.2863}.
\begin{figure}
    \centering
    \includegraphics[width=\columnwidth, clip]{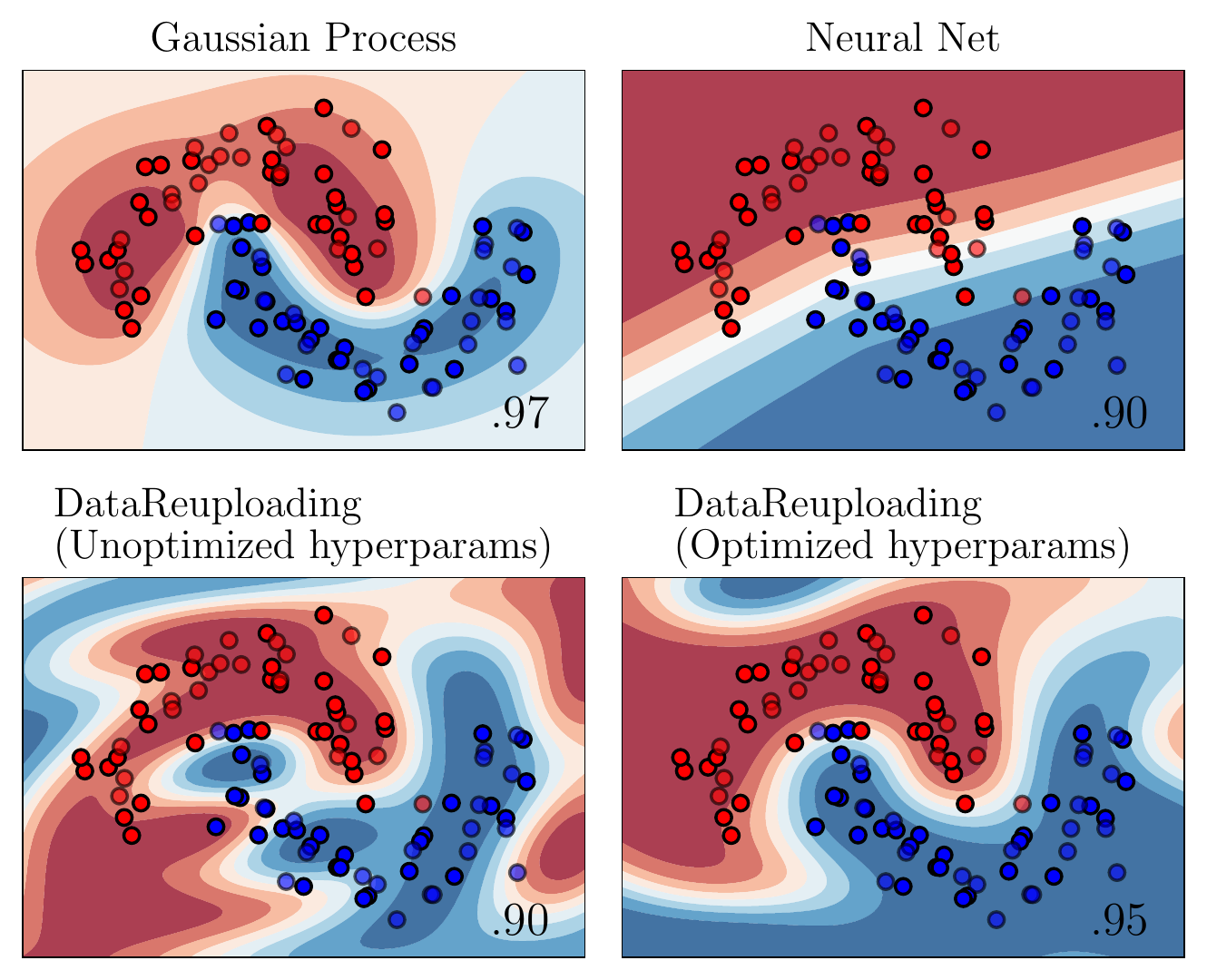}
    \caption{Comparing results of hyperparameter optmization in a quantum classifier against classical methods. The validation score is shown in the inset for each model. In case of the quantum classifier, the decision boundary obtained without hyperparameter optimization has a worse performance compared to optimized hyperparameters. The hyperparameter optimization is faster since we can use a gradient-based optimization by taking implicit gradients of the validation loss.}
    \label{fig:decision}
\end{figure}
\subsection{Hyperparameter optimization in a quantum classifier}
\label{sec:hyperparameter}
We now consider an example motivated by the applications of implicit differentiation to hyperparameter optimization in machine learning~\cite{blondelEfficientModularImplicit2022,jiBilevelOptimizationMachine2021}. Hyperparameters play an important role in model selection and controlling the complexity of machine learning models~\cite{Lever2016}. Our demonstration uses the data-reuploading classifier proposed in \cite{perez-salinasDataReuploadingUniversal2020}, where a single-qubit quantum system is shown to be a universal classifier similar to neural networks with at least one hidden layer. In addition to training the  parameters of the classifier, we consider adding hyperparameters and tuning them through gradient descent, using so-called hyper-gradients~\cite{maclaurinGradientbasedHyperparameterOptimization2015} calculated by implicit differentiation.

A set of inputs $(\alpha_i, \beta_i)$, each with class labels $c_i \in \{0, 1\}$, forms our data set. The inputs are normalized and loaded into a quantum circuit as the parameters $(\alpha_i, \beta_i, 0)$ of a single-qubit unitary given by $U(x_i)$ ({$x_i\equiv(\alpha_i, \beta_i, 0)$}) followed by unitaries $U(z_l)$ where $z_l$ represents model parameters for a layer $l$ of quantum gates. The unitaries are simply rotations on the Bloch sphere with three parameters and we set the last parameter to zero. After applying $L$ layers of such unitaries on some initial quantum state $\ket{\psi_0}$ we obtain the variational state with trainable parameters $z \equiv \{z_L, \dots, z_0\}$ as
\begin{equation}
\ket{\psi(z)} = U(z_L)U(x_i)\dots U(z_0)U(x_i) \ket{\psi_0}.
\end{equation}

Therefore we interleave unitaries that load the data and unitaries that act as learnable weights. A measurement of the operator $\sigma_z$ on the output state gives a single-shot outcome representing our prediction for the class label $c_i$. We take a simpler approach than \cite{perez-salinasDataReuploadingUniversal2020} for the measurements and trainable loss function. By repeating the measurement, we can obtain expectation values of the measurement operator. After normalizing the expectation values to the range $[0, 1]$ such that they can be interpreted as probabilities $p_i$, we define the binary cross-entropy loss function on the training data as
\begin{equation}
    \mathcal L_{\text{bincross}}(z) = \frac{1}{N}\sum_i c_i \log p_i + (1 - c_i)\log(1-p_i).
\end{equation}
Now we can introduce additional hyperparameters in the loss by considering an L2 (ridge regularization~\cite{10.5555/1162264}) penalty on the parameters of each layer to define the training loss
\begin{equation}
\label{eq:loss-hyperparam}
    \mathcal L_{T}(z, a) = \mathcal L_{\text{bincross}}(z) + \sum_l a_l ||z_l||_2,
\end{equation}
where $a = \{a_l\}$ are the hyperparameters for the parameters $z_l$ in each layer of the variational quantum circuit. We optimize the parameters $z$ to minimize the training loss as
\begin{equation}
    z^*(a) = \arg \, \min_{z} L_{T}(z, a).
\end{equation}
A validation loss is constructed using a validation set of the data not used for training 
\begin{equation}
    L_{V}(a) = L_{\text{bincross}}(z^{*}(a)).
\end{equation}
We can minimize the validation loss with respect to the hyperparameters $a$ using gradient descent and the learned parameters $z^*(a)$. The gradient computation $\partial_a L_{V}(a)$ requires differentiating the function $z^*(a)$, which we evaluate through implicit differentiation.
\begin{figure}[htb]
\centering
    \includegraphics[width=0.85\columnwidth]{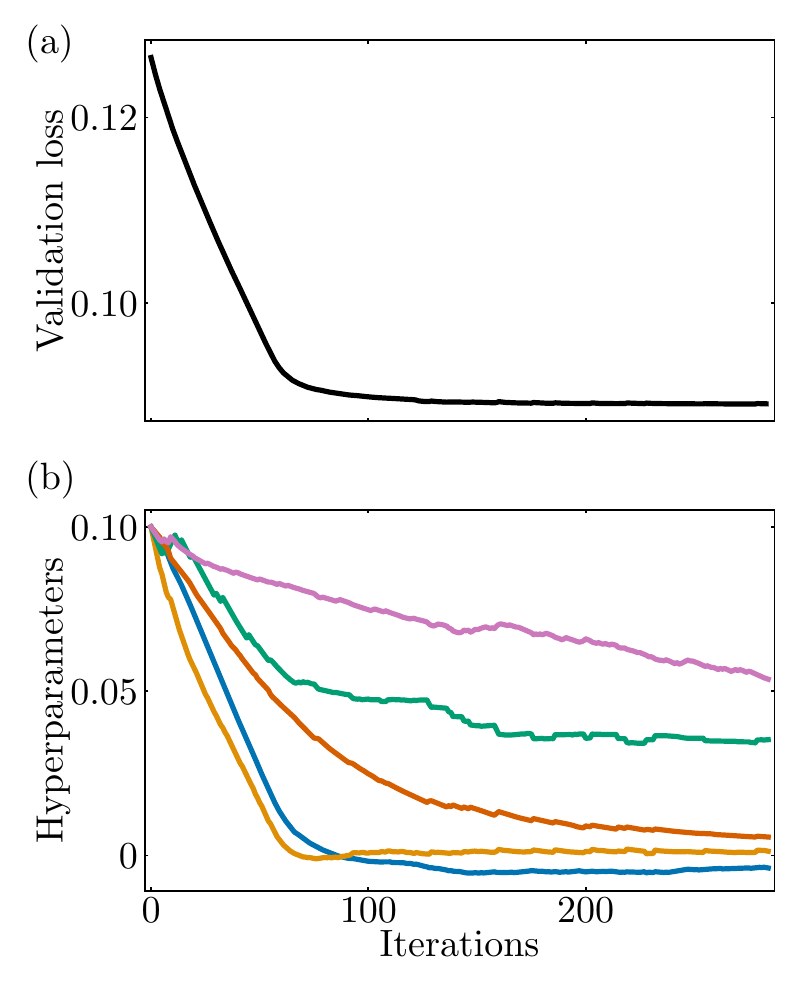}
    \caption{Hyperparameter optmization using implicit gradients. (a) The validation loss as a function of gradient-based updates of the hyperparameters. We compute the gradients of five hyperparameters (one for each layer in the circuit) using implicit differentiation of the validation loss in a quantum classifier. (b) The value of the five hyperparameters as we optimize them by minimizing the validation loss. The implicitly computed gradients allow a meta optimization of the hyperparameters going beyond a costly grid search in the hyperparameter space.}
    \label{fig:hyperparam-grad}
\end{figure}
We consider a single-qubit data-reuploading classifier with $L=5$ layers and five hyperparameters, corresponding to each $a_l$ controlling the strength of the L2 regularization in \eqref{eq:loss-hyperparam}. In \figref{fig:hyperparam-grad}, we show how the validation loss and the hyperparameters change with each gradient-descent update using implicit differentiation to compute the so-called hypergradients. We use the stochastic gradient-descent algorithm for both parameter and hyperparameter optimization with a learning rate of $0.01$. All the hyperparameters are set to the same numerical value at the start of their optimization. As the validation loss improves, we find that the optimal hyperparameters differ significantly and converge to different values.
\begin{figure}
    \centering
    \includegraphics[width=\columnwidth]{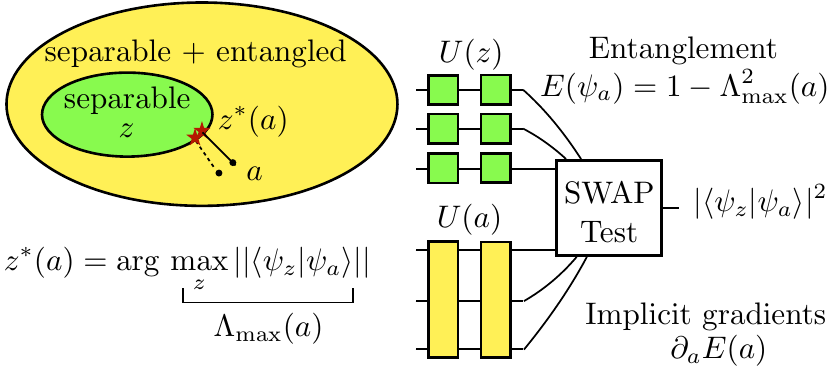}
    \caption{Variational generation of entangled quantum states using implicit differentiation. The distance of a variational quantum state $\ket{\psi_a}$ to the closest separable state $\ket{\psi_{z^{*}(a)}}$ can define a geometric measure of entanglement $E(\psi_a)$. This measure is dependent on the variational parameters $a$. The closest separable state can be obtained by maximizing $|\expec{\psi_z|\psi_a}|^2$ computed using a SWAP test. The optimization implicitly defines the closest separable state $z^{*}(a)$ and therefore the measure of entanglement, $E(a)$. The gradient of the entanglement measure $\partial_a E(a)$ can be computed using implicit differentiation to maximize the entanglement using gradient-based optimization.}
    \label{fig:entanglement-description}
\end{figure}
The effect of the hyperparameter optimization on the classifier performance can be further seen by considering the decision boundaries in \figref{fig:decision} where we also show a comparison to classical machine learning models such as Gaussian processes or simple neural networks. We can see that the decision boundary of the quantum classifier with an unoptimized set of hyperparameters seems to give a worse performance on a validation set compared to the case where hyperparameters are optimized.

The behaviour of the data-reuploading classifier with and without hyperparameter optimization shows that tuning the hyperparameters leads to a better learning of the decision boundary. In most hyperparameter tuning techniques, a grid search is performed for each of the hyperparameters. However, with implicit differentiation, it is possible to optimize millions of hyperparameters faster~\cite{lorraineOptimizingMillionsHyperparameters2020}.
\subsection{Variational optimization of a geometric measure of quantum entanglement}
\label{sec:entanglement}
Quantum entanglement is an important resource for quantum communication, quantum cryptography, and plays a pivotal role in the possible speedup provided by quantum algorithms~\cite{jozsaRoleEntanglementQuantumcomputational2003}. An entangled quantum state can be defined as a state that cannot be written as a separable state, i.e., as $\ket{\psi} = \ket{\psi_1} \otimes \ket{\psi_2} \otimes \ket{\psi_3} \otimes \cdots \otimes \ket{\psi_N}$. A measure of entanglement for an arbitrary quantum state is any positive real-valued function of the state that cannot increase under local operations on parts of the state and classical communication. There exists several examples of quantum entanglement measures, e.g., concurrence, logarithmic negativity, von Neumann entropy~\cite{Horodecki2008}. A number of such measures become simple for bipartite quantum systems, but quantifying multiparitite entanglement is generally difficult. We use the geometric measure of quantum entanglement defined in~\cite{Wei2003} as,
\begin{equation}
\label{eq:entanglement-solution}
    E(\psi) = 1 - \max_{\ket{\psi'}} |\expec{\psi'|\psi}|^2,
\end{equation}
where the optimization is over the set of separable states $\ket{\psi'}$. This geometric measure of entanglement can be intuitively understood as the distance between an arbitrary quantum state and the nearest separable quantum state found by solving \eqref{eq:entanglement-solution}. We now demonstrate how a variational algorithm can generate entangled quantum states by optimizing this geometric measure using implicit differentiation.

\begin{figure}
\centering
    \includegraphics[width=0.95\columnwidth]{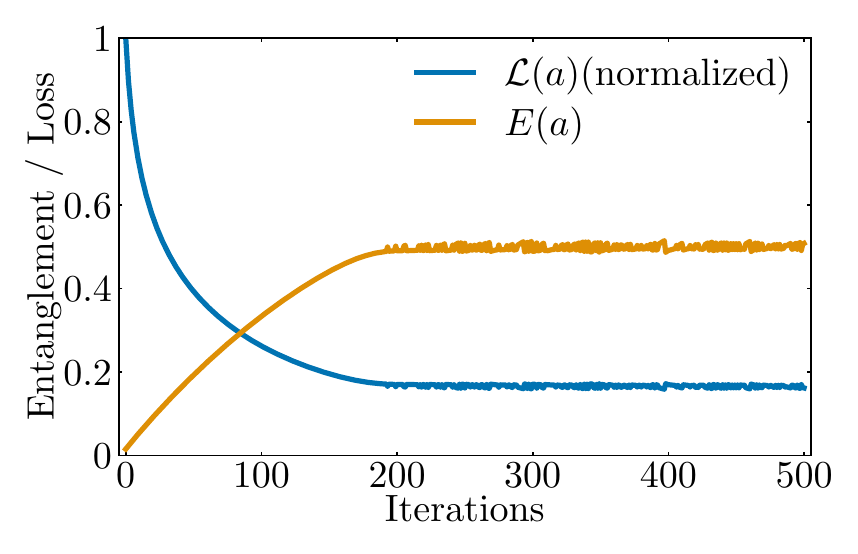}
    \caption{\label{fig:entanglement-results}Optimization of a geometric entanglement measure using gradient descent. The loss and entanglement measure after each update of the variational parameters $a$. Using implicit differentiation, we compute the gradient of the loss function and minimize it to maximize the entanglement measure.}
\end{figure}

\begin{figure}[htp]
\centering
    \includegraphics[width=0.95\columnwidth]{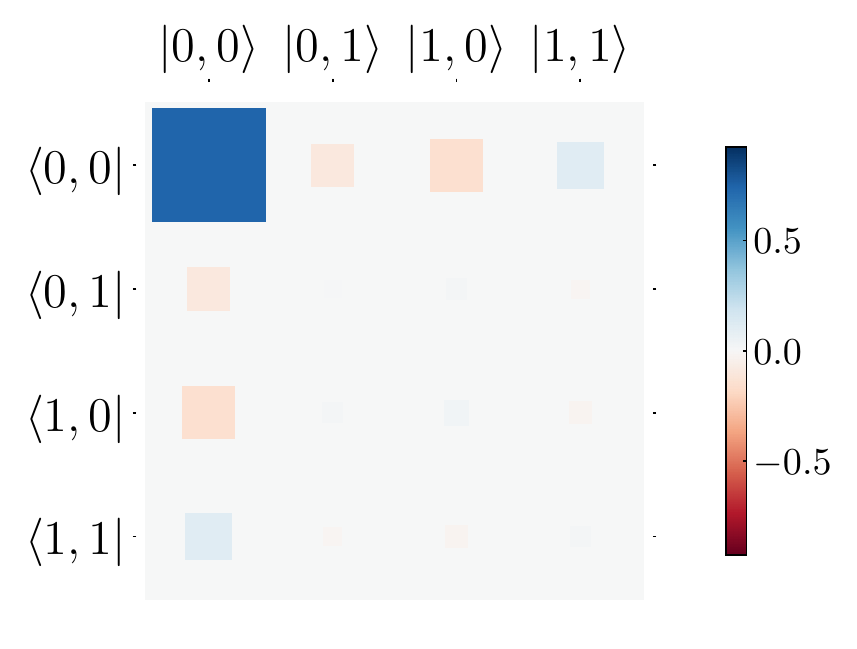}
    \includegraphics[width=0.95\columnwidth]{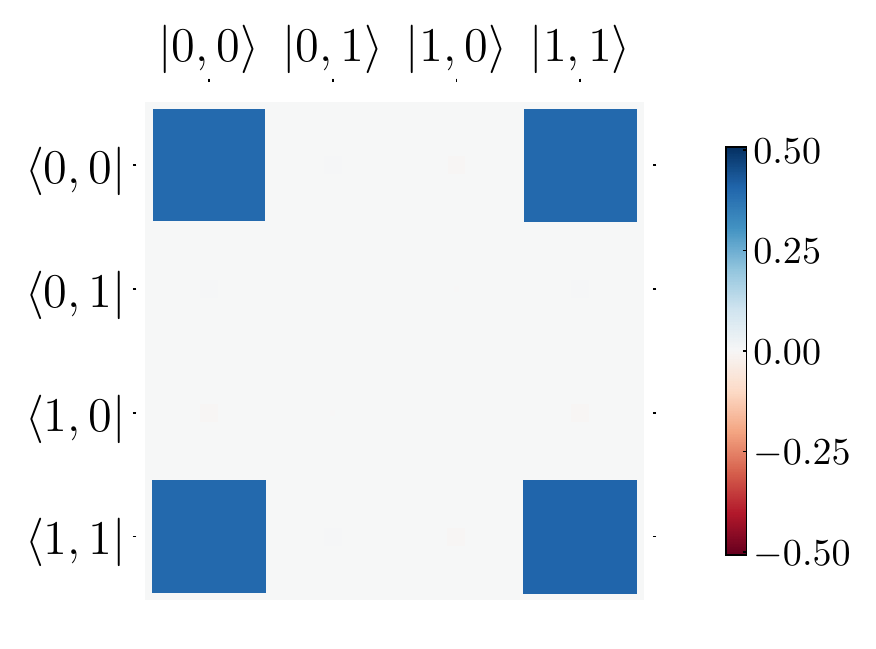}
    \caption{\label{fig:bell}The final state $\ket{\psi_a}$ (bottom) generated by our variational approach starting from a random set of initial parameters (top). We use a variational quantum circuit that can create entanglement but learn the parameters using implicit differentiation and gradient descent; see \figref{fig:entanglement-description}. The geometric measure of entanglement uses another variational quantum circuit that only consists of single qubit operations, i.e., can only create separable states, such that computing an overlap between the separable state and a (possibly) entangled state gives a measure of entanglement. Implicit differentiation allows us to compute gradients through this geometric measure of entanglement. Optimizing the entanglement using gradient descent, we arrive at a maximally entangled Bell state $\frac{\ket{00} + \ket{11}}{\sqrt{2}}$ without ever explicitly defining the Bell state.}
\end{figure}

In \figref{fig:entanglement-description} we show how two different variational quantum circuits can be used to create an entangled state $\ket{\psi_a}$ with variational parameters $a$. One circuit consists of only local (single-qubit) unitaries specified by parameters $z$, which creates separable states $\ket{\psi_z}$. The other circuit contains general operations that could possibly create entangled states $\ket{\psi_a}$. A SWAP test, its improved version~\cite{cincioLearningQuantumAlgorithm2018}, or any choice of randomized measurement overlap estimator~\cite{PhysRevLett.124.010504,gueriniQuasiprobabilisticStateoverlapEstimator2021,elbenRandomizedMeasurementToolbox2022}, can provide a measure of the distinguishability between the two states~\cite{barenco1997stabilization, Burhman2001}, and we can obtain a separable quantum state $\ket{\psi_{z^{*}(a)}}$ that is close to a fixed state $\ket{\psi_a}$ by the following optimization
\begin{equation}
\label{eq:entanglement-z}
    z^{*}(a) = \arg\,\max_{z} {\vert\expec{\psi_z|\psi_a}\vert}^2.
\end{equation}
The optimization problem defines an implicit function $z^{*}(a)$ that gives a measure of the entanglement of the state $E(\ket{\psi_a}) = E(a)$. Practically, we solve \eqref{eq:entanglement-z} by minimizing $-\vert\expec{\psi_z|\psi_a}\vert^2$. Now, using implicit differentiation we can obtain the gradients of the entanglement measure $\partial_a E(a)$ and maximize it. In \figref{fig:entanglement-results}, we show the results of this procedure by minimizing the loss function
\begin{equation}
\label{eq:entanglement-a}
    \mathcal L(a) = -\log{E(a)}
\end{equation}
The gradient of the loss function can be computed without the need for backpropagation through the quantum algorithm that generates the nearest separable state. We only need the value of the overlap and local gradients at $z^{*}(a)$. We optimize both \eqref{eq:entanglement-z} and \eqref{eq:entanglement-a} with stochastic gradient-descent with a learning rate of $0.001$.
The results of this bi-level optimization is shown in \figref{fig:entanglement-results} where we can see that the entanglement measure increases as the loss function (negative log of the entanglement) decreases. 
As a result we variationally obtained one of the Bell states starting from a random initial state, without ever defining it, as shown in \figref{fig:bell}. We can similarly extend this approach to multiple qubits to generate entangled quantum states variationally using gradient-based optimization. The algorithm is sensitive to the choice of the loss function, optimization, and hyperparameters and here we demonstrate one choice that allowed the creation of Bell states. In future investigations, we can determine the best choice of loss function and hyperparameters for a robust generation of entangled quantum states. 
\section{Conclusion}
\label{sec:conclusion}
In this work, we have presented a common approach to applying implicit differentiation to quantum algorithms that encompasses a range of use-cases—from calculating ground-state property gradients using a variational quantum algorithm to hyperparameter optimization in quantum machine learning. We show how leveraging modern automatic-differentiation tools working in tandem with quantum software can simplify the application of implicit differentiation  to wide variety of problems.

Our work shows how we can go beyond the Hellman-Feynmann theorem to compute derivatives of the expectation of arbitrary operators and not just the total energy in a parameterized Hamiltonian. As a demonstration, we take a simple spin-chain model to show how a variational quantum algorithm can be used to obtain ground-state  gradients by a parameterized quantum circuit. Such calculations may be easily performed on emerging NISQ devices.

Finally, we demonstrated two powerful applications of implicit differentiation: hyperparameter optimization of quantum machine learning algorithms, and variationally generating entangled quantum states using a geometric measure of entanglement. We demonstrate how hyperparameter optimization with implicit differentiation can not only be applied in the classical machine learning setting but also in quantum algorithms. We have shown that optimizing regularization strength using implicit differentiation could lead to better performance where implicit differentiation allowed a meta-optimization on the hyperparameters to avoid a costly grid search. The idea of entanglement generation using a bi-level optimization in a variational setting could possibly allow creating large multipartite entanglement beyond specific classes such as GHZ or W states.

Implicit differentiation is an idea as old as calculus itself with the first notable application by Fermat to compute tangents in the folium of Descartes curve~\cite{paradis2004fermat}. As it was with the idea of applying gradient descent in both classical and quantum machine learning, new breakthroughs came with efficient and algorithmic ways to compute gradients by automatic differentiation. Recent demonstrations of implicit differentiation for inverse design in steady-state quantum systems~\cite{vargas-hernandezFullyDifferentiableOptimization2021} show how powerful this idea is with potential applications to quantum chemistry, control, and quantum machine learning. Our work opens up avenues to explore ideas such as neural quantum ODEs and compute interesting physical quantities using quantum computers and implicit differentiation.

The practical implementations of implicit differentiation on quantum hardware would require access to gradient computations as well as fast solutions to the optimality condition. Future works could investigate in detail how feasible such computations are when scaled to larger systems. It is also important to analyze how errors affect the gradient computation and its scaling. During the preparation of this manuscript, we became aware of another work that focuses on using implicit differentiation to compute the fidelity susceptibility and analyzes some of these issues~\cite{DiMatteo2022}. We hope that our work together with similar demonstrations of implicit differentiation~\cite{vargas-hernandezFullyDifferentiableOptimization2021} can find many more applications in the emerging area of quantum computation and quantum machine learning.

\section*{Acknowledgements}
 SA acknowledges discussions with Roeland Wiersema and Josh Izaac. JFCA acknowledges discussions with Federico Becca and Sandro Sorella. This work was supported by Mitacs through the Mitacs Accelerate program. JFCA acknowledges support from the Natural Sciences and Engineering Research Council (NSERC), the Shared Hierarchical Academic Research Computing Network (SHARCNET), Compute Canada, and the Canadian Institute for Advanced Research (CIFAR) AI chair program. Resources used in preparing this research were provided, in part, by the Province of Ontario, the Government of Canada through CIFAR, and companies sponsoring the Vector Institute 
 \url{www.vectorinstitute.ai/#partners}
\bibliographystyle{unsrt}
\bibliography{references}

\begin{thebibliography}{10}

\bibitem{Cerezo2021}
M.~Cerezo, Andrew Arrasmith, Ryan Babbush, Simon~C. Benjamin, Suguru Endo,
  Keisuke Fujii, Jarrod~R. McClean, Kosuke Mitarai, Xiao Yuan, Lukasz Cincio,
  and Patrick~J. Coles.
\newblock {Variational quantum algorithms}.
\newblock {\em Nat. Rev. Phys.}, 3(9):625--644, 2021.

\bibitem{RevModPhys.94.015004}
Kishor Bharti, Alba Cervera-Lierta, Thi~Ha Kyaw, Tobias Haug, Sumner
  Alperin-Lea, Abhinav Anand, Matthias Degroote, Hermanni Heimonen, Jakob~S.
  Kottmann, Tim Menke, Wai-Keong Mok, Sukin Sim, Leong-Chuan Kwek, and Al\'an
  Aspuru-Guzik.
\newblock Noisy intermediate-scale quantum algorithms.
\newblock {\em Rev. Mod. Phys.}, 94:015004, 2022.

\bibitem{PhysRevResearch.2.033125}
Yuxuan Du, Min-Hsiu Hsieh, Tongliang Liu, and Dacheng Tao.
\newblock Expressive power of parametrized quantum circuits.
\newblock {\em Phys. Rev. Research}, 2:033125, 2020.

\bibitem{elbenRandomizedMeasurementToolbox2022}
Andreas Elben, Steven~T. Flammia, Hsin-Yuan Huang, Richard Kueng, John
  Preskill, Beno{\^{i}}t Vermersch, and Peter Zoller.
\newblock {The randomized measurement toolbox}.
\newblock (arXiv:2203.11374), 2022.

\bibitem{obrienCalculatingEnergyDerivatives2019}
Thomas~E. O'Brien, Bruno Senjean, Ramiro Sagastizabal, Xavier {Bonet-Monroig},
  Alicja Dutkiewicz, Francesco Buda, Leonardo DiCarlo, and Lucas Visscher.
\newblock Calculating energy derivatives for quantum chemistry on a quantum
  computer.
\newblock {\em Npj Quantum Inf.}, 5(1):1--12, 2019.

\bibitem{pulayInitioCalculationForce1969}
P.~Pulay.
\newblock Ab initio calculation of force constants and equilibrium geometries
  in polyatomic molecules.
\newblock {\em Mol. Phys.}, 17(2):197--204, 1969.

\bibitem{PhysRevResearch.2.013129}
Kosuke Mitarai, Yuya~O. Nakagawa, and Wataru Mizukami.
\newblock Theory of analytical energy derivatives for the variational quantum
  eigensolver.
\newblock {\em Phys. Rev. Research}, 2:013129, 2020.

\bibitem{kassalQuantumAlgorithmMolecular2009}
Ivan Kassal and Al{\'a}n {Aspuru-Guzik}.
\newblock Quantum algorithm for molecular properties and geometry optimization.
\newblock {\em J. Chem. Phys.}, 131(22):224102, 2009.

\bibitem{sandvikComputationalStudiesQuantum2010a}
Anders~W. Sandvik.
\newblock Computational {{Studies}} of {{Quantum Spin Systems}}.
\newblock {\em AIP Conf. Proc.}, 1297(1):135--338, 2010.

\bibitem{PhysRevE.76.022101}
Wen-Long You, Ying-Wai Li, and Shi-Jian Gu.
\newblock Fidelity, dynamic structure factor, and susceptibility in critical
  phenomena.
\newblock {\em Phys. Rev. E}, 76:022101, 2007.

\bibitem{PhysRevLett.99.095701}
Lorenzo Campos~Venuti and Paolo Zanardi.
\newblock Quantum critical scaling of the geometric tensors.
\newblock {\em Phys. Rev. Lett.}, 99:095701, 2007.

\bibitem{bergholmPennyLaneAutomaticDifferentiation2022}
Ville Bergholm, Josh Izaac, Maria Schuld, Christian Gogolin, Shahnawaz Ahmed,
  Vishnu Ajith, M.~Sohaib Alam, et~al.
\newblock {{PennyLane}}: {{Automatic}} differentiation of hybrid
  quantum-classical computations.
\newblock (arXiv:1811.04968), 2022.

\bibitem{broughtonTensorFlowQuantumSoftware2021}
Michael Broughton, Guillaume Verdon, Trevor McCourt, Antonio~J. Martinez,
  Jae~Hyeon Yoo, Sergei~V. Isakov, Philip Massey, Ramin Halavati,
  Murphy~Yuezhen Niu, Alexander Zlokapa, Evan Peters, Owen Lockwood, Andrea
  Skolik, Sofiene Jerbi, Vedran Dunjko, Martin Leib, Michael Streif, David
  Von~Dollen, Hongxiang Chen, Shuxiang Cao, Roeland Wiersema, Hsin-Yuan Huang,
  Jarrod~R. McClean, Ryan Babbush, Sergio Boixo, Dave Bacon, Alan~K. Ho,
  Hartmut Neven, and Masoud Mohseni.
\newblock {{TensorFlow Quantum}}: {{A Software Framework}} for {{Quantum
  Machine Learning}}.
\newblock (arXiv:2003.02989), 2021.

\bibitem{PhysRevA.98.032309}
K.~Mitarai, M.~Negoro, M.~Kitagawa, and K.~Fujii.
\newblock Quantum circuit learning.
\newblock {\em Phys. Rev. A}, 98:032309, 2018.

\bibitem{wierichsGeneralParametershiftRules2022}
David Wierichs, Josh Izaac, Cody Wang, and Cedric Yen-Yu Lin.
\newblock General parameter-shift rules for quantum gradients.
\newblock {\em Quantum}, 6:677, 2022.

\bibitem{blondelEfficientModularImplicit2022}
Mathieu Blondel, Quentin Berthet, Marco Cuturi, Roy Frostig, Stephan Hoyer,
  Felipe {Llinares-L{\'o}pez}, Fabian Pedregosa, and Jean-Philippe Vert.
\newblock Efficient and {{Modular Implicit Differentiation}}.
\newblock (arXiv:2105.15183), 2021.

\bibitem{code}
Code and data for the article "Implicit differentiation of variational quantum
  algorithms" is available at
  \href{https://github.com/quantshah/quantum-implicit-differentiation}{https://github.com/quantshah/quantum-implicit-differentiation}.

\bibitem{krantz2002implicit}
Steven~George Krantz and Harold~R Parks.
\newblock {\em The implicit function theorem: history, theory, and
  applications}.
\newblock Springer Science \& Business Media, 2002.

\bibitem{Hormander1973}
Lars Hormander.
\newblock {\em An introduction to complex analysis in several variables}.
\newblock Elsevier, 1973.

\bibitem{chang2003analytic}
Hung-Chieh Chang, Wei He, and Nagabhushana Prabhu.
\newblock The analytic domain in the implicit function theorem.
\newblock {\em JIPAM. J. Inequal. Pure Appl. Math}, 4(1), 2003.

\bibitem{lorraineOptimizingMillionsHyperparameters2020}
Jonathan Lorraine, Paul Vicol, and David Duvenaud.
\newblock Optimizing {{Millions}} of {{Hyperparameters}} by {{Implicit
  Differentiation}}.
\newblock In {\em Proceedings of the {{Twenty Third International Conference}}
  on {{Artificial Intelligence}} and {{Statistics}}}, pages 1540--1552. {PMLR},
  2020.

\bibitem{Rajeswaran2019}
Aravind Rajeswaran, Sham~M. Kakade, Chelsea Finn, and Sergey Levine.
\newblock {Meta-learning with implicit gradients}.
\newblock {\em Advances in {Neural Information Processing Systems}}, 32, 2019.

\bibitem{PhysRev.56.340}
R.~P. Feynman.
\newblock Forces in molecules.
\newblock {\em Phys. Rev.}, 56:340--343, 1939.

\bibitem{schuld2019evaluating}
Maria Schuld, Ville Bergholm, Christian Gogolin, Josh Izaac, and Nathan
  Killoran.
\newblock Evaluating analytic gradients on quantum hardware.
\newblock {\em Phys. Rev. A.}, 99(3):032331, 2019.

\bibitem{PhysRevA.103.012405}
Andrea Mari, Thomas~R. Bromley, and Nathan Killoran.
\newblock Estimating the gradient and higher-order derivatives on quantum
  hardware.
\newblock {\em Phys. Rev. A}, 103:012405, 2021.

\bibitem{PhysRevB.99.165130}
Gergely Barcza, Florian Gebhard, Thorben Linneweber, and \"Ors Legeza.
\newblock Ground-state properties of the symmetric single-impurity anderson
  model on a ring from density-matrix renormalization group, hartree-fock, and
  gutzwiller theory.
\newblock {\em Phys. Rev. B}, 99:165130, 2019.

\bibitem{PhysRevB.76.180403}
Min-Fong Yang.
\newblock Ground-state fidelity in one-dimensional gapless models.
\newblock {\em Phys. Rev. B}, 76:180403, 2007.

\bibitem{kirillovExactSolutionHeisenberg1986}
A.~N. Kirillov and N.~Yu. Reshetikhin.
\newblock Exact solution of the {{Heisenberg XXZ}} model of spin s.
\newblock {\em {J. Sov. Math.}}, 35(4):2627--2643, 1986.

\bibitem{PhysRevA.87.043606}
Juan Carrasquilla, Salvatore~R. Manmana, and Marcos Rigol.
\newblock Scaling of the gap, fidelity susceptibility, and bloch oscillations
  across the superfluid-to-mott-insulator transition in the one-dimensional
  bose-hubbard model.
\newblock {\em Phys. Rev. A}, 87:043606, 2013.

\bibitem{sakaiQuantumSpinFluid2018}
T{\^o}ru Sakai and Hiroki Nakano.
\newblock Quantum {{Spin Fluid Behaviors}} of the {{Kagome-}} and
  {{Triangular-Lattice Antiferromagnets}}.
\newblock {\em J. Phys. Conf. Ser.}, 969(1):012127, 2018.

\bibitem{boeddekerComputationComplexvaluedGradients2019}
Christoph Boeddeker, Patrick Hanebrink, Lukas Drude, Jahn Heymann, and Reinhold
  {Haeb-Umbach}.
\newblock On the {{Computation}} of {{Complex-valued Gradients}} with
  {{Application}} to {{Statistically Optimum Beamforming}}.
\newblock (arXiv:1701.00392), 2017.

\bibitem{PhysRevB.101.245139}
Hao Xie, Jin-Guo Liu, and Lei Wang.
\newblock Automatic differentiation of dominant eigensolver and its
  applications in quantum physics.
\newblock {\em Phys. Rev. B}, 101:245139, 2020.

\bibitem{jax2018github}
James Bradbury, Roy Frostig, Peter Hawkins, Matthew~James Johnson, Chris Leary,
  Dougal Maclaurin, George Necula, Adam Paszke, Jake Vander{P}las, Skye
  Wanderman-{M}ilne, and Qiao Zhang.
\newblock {JAX}: composable transformations of {P}ython+{N}um{P}y programs,
  2018.

\bibitem{carleoSolvingQuantumManybody2017}
Giuseppe Carleo and Matthias Troyer.
\newblock Solving the quantum many-body problem with artificial neural
  networks.
\newblock {\em Science}, 355(6325):602--606, 2017.

\bibitem{PhysRevLett.69.2863}
Steven~R. White.
\newblock Density matrix formulation for quantum renormalization groups.
\newblock {\em Phys. Rev. Lett.}, 69:2863--2866, 1992.

\bibitem{jiBilevelOptimizationMachine2021}
Kaiyi Ji.
\newblock Bilevel {{Optimization}} for {{Machine Learning}}: {{Algorithm
  Design}} and {{Convergence Analysis}}.
\newblock (arXiv:2108.00330), 2021.

\bibitem{Lever2016}
Jake Lever, Martin Krzywinski, and Naomi Altman.
\newblock {Points of Significance: Regularization}.
\newblock {\em {Nat. Methods}}, 13(10):803--804, 2016.

\bibitem{perez-salinasDataReuploadingUniversal2020}
Adri{\'a}n {P{\'e}rez-Salinas}, Alba {Cervera-Lierta}, Elies {Gil-Fuster}, and
  Jos{\'e}~I. Latorre.
\newblock Data re-uploading for a universal quantum classifier.
\newblock {\em Quantum}, 4:226, 2020.

\bibitem{maclaurinGradientbasedHyperparameterOptimization2015}
Dougal Maclaurin, David Duvenaud, and Ryan Adams.
\newblock Gradient-based {{Hyperparameter Optimization}} through {{Reversible
  Learning}}.
\newblock In {\em Proceedings of the 32nd {{International Conference}} on
  {{Machine Learning}}}, pages 2113--2122. {PMLR}, 2015.

\bibitem{10.5555/1162264}
Christopher~M. Bishop.
\newblock {\em Pattern Recognition and Machine Learning (Information Science
  and Statistics)}.
\newblock Springer-Verlag, Berlin, Heidelberg, 2006.

\bibitem{jozsaRoleEntanglementQuantumcomputational2003}
Richard Jozsa and Noah Linden.
\newblock On the role of entanglement in quantum-computational speed-up.
\newblock {\em Proceedings of the Royal Society of London. Series A:
  Mathematical, Physical and Engineering Sciences}, 459(2036):2011--2032, 2003.

\bibitem{Horodecki2008}
Ryszard Horodecki, Pawe\l{} Horodecki, Micha\l{} Horodecki, and Karol
  Horodecki.
\newblock Quantum entanglement.
\newblock {\em Rev. Mod. Phys.}, 81:865--942, 2009.

\bibitem{Wei2003}
Tzu-Chieh Wei and Paul~M. Goldbart.
\newblock Geometric measure of entanglement and applications to bipartite and
  multipartite quantum states.
\newblock {\em Phys. Rev. A}, 68:042307, 2003.

\bibitem{cincioLearningQuantumAlgorithm2018}
Lukasz Cincio, Yi{\u g}it Suba{\c s}{\i}, Andrew~T. Sornborger, and Patrick~J.
  Coles.
\newblock Learning the quantum algorithm for state overlap.
\newblock {\em New Journal of Physics}, 20(11):113022, November 2018.

\bibitem{PhysRevLett.124.010504}
Andreas Elben, Beno\^{\i}t Vermersch, Rick van Bijnen, Christian Kokail, Tiff
  Brydges, Christine Maier, Manoj~K. Joshi, Rainer Blatt, Christian~F. Roos,
  and Peter Zoller.
\newblock Cross-platform verification of intermediate scale quantum devices.
\newblock {\em Phys. Rev. Lett.}, 124:010504, Jan 2020.

\bibitem{gueriniQuasiprobabilisticStateoverlapEstimator2021}
Leonardo Guerini, Roeland Wiersema, Juan~Felipe Carrasquilla, and Leandro
  Aolita.
\newblock Quasiprobabilistic state-overlap estimator for {{NISQ}} devices,
  December 2021.

\bibitem{barenco1997stabilization}
Adriano Barenco, Andre Berthiaume, David Deutsch, Artur Ekert, Richard Jozsa,
  and Chiara Macchiavello.
\newblock Stabilization of quantum computations by symmetrization.
\newblock {\em SIAM Journal on Computing}, 26(5):1541--1557, 1997.

\bibitem{Burhman2001}
Harry Buhrman, Richard Cleve, John Watrous, and Ronald de~Wolf.
\newblock Quantum fingerprinting.
\newblock {\em Phys. Rev. Lett.}, 87:167902, 2001.

\bibitem{paradis2004fermat}
Jaume Parad{\'\i}s, Josep Pla, and Pelegr{\'\i} Viader.
\newblock Fermat and the quadrature of the folium of descartes.
\newblock {\em The American Mathematical Monthly}, 111(3):216--229, 2004.

\bibitem{vargas-hernandezFullyDifferentiableOptimization2021}
Rodrigo~A. {Vargas-Hern{\'a}ndez}, Ricky T.~Q. Chen, Kenneth~A. Jung, and Paul
  Brumer.
\newblock Fully differentiable optimization protocols for non-equilibrium
  steady states.
\newblock {\em New J. Phys.}, 23(12):123006, 2021.

\bibitem{DiMatteo2022}
Olivia {Di Matteo} and R.~M. Woloshyn.
\newblock {Quantum computing fidelity susceptibility using automatic
  differentiation}.
\newblock (arXiv:2207.06526), 2022.

\end{thebibliography}
\end{document}